\documentclass[letterpaper,twocolumn,superscriptaddress,showpacs,preprintnumbers,amsmath,amssymb,showkeys]{revtex4}
\usepackage{graphicx}% Include figure files
\usepackage{bm}% bold math
\usepackage[bookmarks=true]{hyperref} 
\usepackage{times}
\usepackage{euler}

\usepackage{color}

\newcommand{\er}[1]{\mathbb{#1}}

\newcommand{\X}{{\bf X}}
\newcommand{\Z}{{\bf Z}}
\newcommand{\Y}{{\bf Y}}
\newcommand{\I}{\openone}
\newcommand{\CZ}{{\tt CSIGN}}%{{\bf Z}^{C}}
\newcommand{\NS}{{\tt NS}}
\newcommand{\CNOT}{{\tt CNOT}}
\newcommand{\ox}{\otimes}
\newcommand{\e}{\epsilon}
\newcommand{\dt}{\delta}
\newcommand{\op}[1]{{\bf #1}}
\newcommand{\ket}[1]{\left|#1\right\rangle}
\newcommand{\ketbra}[2]{\left|#1\rangle\langle#2\right|}
\newcommand{\logic}[1]{\mathfrak{#1}}

\begin{document}
%\title{Erasure Thresholds for Linear Optics Quantum Computing}
\title{Thresholds for Linear Optics Quantum Computing with Photon Loss
at the Detectors}

\author{Marcus \surname{Silva}}
\affiliation{Institute for Quantum Computing,
University of Waterloo, 
200 University Ave. W, 
Waterloo, Ontario,
N2L 3G1, Canada}
\affiliation{Department of Physics,
University of Waterloo, 
200 University Ave. W, 
Waterloo, Ontario,
N2L 3G1, Canada}
\author{Martin \surname{R\"otteler}}
\affiliation{
NEC Laboratories America, Inc.,
4 Independence Way,
Princeton, NJ 08540, U.S.A.
}
\author{Christof \surname{Zalka}}
\affiliation{Institute for Quantum Computing,
University of Waterloo, 
200 University Ave. W, 
Waterloo, Ontario,
N2L 3G1, Canada}
\affiliation{Department of Physics,
University of Waterloo, 
200 University Ave. W, 
Waterloo, Ontario,
N2L 3G1, Canada}

\date{\today}

\begin{abstract}
We calculate the error threshold for the linear optics quantum computing proposal
by Knill, Laflamme and Milburn [Nature {\bf 409}, pp. 46--52 (2001)]
under an error model where photon detectors
have efficiency $<100\%$ but all other components -- such as single photon sources,
beam splitters and phase shifters -- are perfect and introduce no errors. 
We make use of the fact that the error model induced by the lossy hardware is that
of an erasure channel, i.e., the error locations are always known.
Using a method based on a Markov chain description of the error correction procedure, our
calculations show that, with the 7 qubit CSS quantum code, the gate error threshold
for fault tolerant quantum computation is bounded below by a value
between $1.78\%$ and $11.5\%$ depending on the construction of the entangling gates. 
\end{abstract}

\pacs{03.67.Lx}

\keywords{error correction, threshold, linear optics, quantum computing, Markov chains}
\maketitle
%%%%%%%%%%%%%%%%%%%%%%%%%%%%%%%%%%%%%%%%%%%%%%%%%%%%%%%%%%%%%%%%%%%%%%%%%
\section{\label{sec:intro}Introduction}
In Refs.~\onlinecite{klm:2001, klm-thr:2000} it was demonstrated how a
quantum computer could be built using only single photon sources, 
passive linear optics elements, and photon detectors. 
Quantum computing proposals that use photons to encode information are particularly 
interesting because of practical applications to quantum communication over
optical fibers, and the natural resilience of photons to decoherence.
This proposal is also
a conceptual departure from other quantum computing proposals because
it requires post-selection of states in order to overcome the
limitations imposed by the choice of physical resources -- namely, the fact
that one cannot make photons interact using passive linear optics elements.
Moreover, even when considering ideal hardware (i.e. lossless and infinitely precise linear optics elements, 
100\% efficient detectors) one {\em must} use error correction
codes to make the implementation efficient~\cite{klm-thr:2000}. However, photon
detectors are necessary for the qubit measurements used in post-selection and error correction, and
good photon detectors are notoriously hard to build. In this paper, we investigate the maximum error 
rate that a linear optics quantum computer,
as proposed in Ref.~\onlinecite{klm:2001}, can sustain, 
assuming that the only source of hardware imperfections is the
finite photon loss at the photon detectors.

The paper is organized as follows.
In Section~\ref{sec:error-model} we briefly describe the construction of a
probabilistic gate as given
in Ref.~\onlinecite{klm:2001} and emphasize the description of the error
model which arises naturally from considering those gates. In Section
\ref{sec:lossy-error-model} we describe an error model based on the same
gate constructions, but assuming that the single photon detectors have
less than perfect efficiency.
In Section~\ref{sec:error-correction} the error correction code
and the circuits used for error correction are described, along with constraints for 
fault-tolerance. Finally, in Section~\ref{sec:thresholds} the recursion relations for the
error rates at different level of concatenated encoding are given under worst case
assumptions, along with a brief description
of how they were calculated, and the threshold values are stated. 

Throughout the paper we will use the Pauli matrices
$\X=\left(\begin{array}{cc}0 & 1\\ 1 & 0\end{array}\right)$,
$\Y=\left(\begin{array}{cr}0 & -i\\ i & 0\end{array}\right)$,
$\Z=\left(\begin{array}{cr}1 & 0\\ 0 & -1\end{array}\right)$ along with the
identity matrix $\openone$.

%%%%%%%%%%%%%%%%%%%%%%%%%%%%%%%%%%%%%%%%%%%%%%%%%%%%%%%%%%%%%%%%%%%%%%%%%
\section{\label{sec:error-model}Ideal Hardware Error Model}

In the efficient linear optics quantum computing proposal put forward
by Knill, Laflamme and Milburn~\cite{klm:2001}, qubits are encoded as a single
photon in one of two optical modes, that is,
$\ket{\logic{0}}$ is represented by the photon number state
$\ket{0}\ket{1}$, and 
$\ket{\logic{1}}$ is represented by the photon number state $\ket{1}\ket{0}$
-- this encoding is also called the {\em dual-rail encoding}~\footnote{In 
a photon number state, each ket represents an optical mode, and the number 
represents the number of photons occupying that mode.}. The only
resources available are single photon sources, passive linear optics elements (such 
as beam splitters and phase shifters), and photon detectors. In our model, the only source
of hardware imperfection is the efficiency of the photon detectors, that is, the photon
sources and passive linear optics elements are assumed to be perfect.
We also assume that 
classical computation and control are delay and error free, and that all sources
of failure -- teleportation and measurement failures -- are statistically independent.

\begin{figure}
\includegraphics[width=7.5cm]{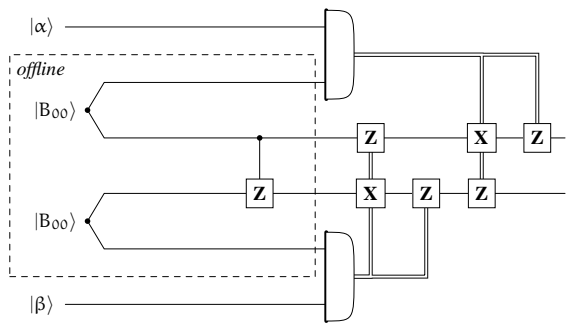}
\caption{\label{fig:c-sign-tele}Gate teleportation protocol for a $\CZ$. 
The resource state is $\ket{B_{00}}=\frac{\ket{\logic{00}}+\ket{\logic{11}}}{\sqrt{2}}$ and
the two qubit measurement is a Bell basis measurement. The double lines indicate operations
conditioned on the outcome of the measurement.
}
\end{figure}

While single qubit operations can be efficiently performed using only phase
shifters and beam-splitters~\cite{zeilinger:1994} -- and therefore can be considered error
free in our model -- two qubit
operations require state post-selection through measurement of ancillary 
modes. If the desired measurement is not obtained, the operation may or may not
have been applied properly, but if the desired measurement is obtained,
the proper operation is guaranteed to have been applied. In Ref.~\onlinecite{klm:2001} a probabilistic
sign shift gate, which performs the operation 
$\NS_-=\ketbra{0}{0}+\ketbra{1}{1}-\ketbra{2}{2}$ on number states of a given mode, is described.
This gate succeeds with probability $\frac{1}{4}$. A probabilistic entangling gate
$\CZ$, which performs the operation 
$\ketbra{\logic{00}}{\logic{00}}+\ketbra{\logic{01}}{\logic{01}}+\ketbra{\logic{10}}{\logic{10}}-\ketbra{\logic{11}}{\logic{11}}$ on two qubits, 
can be constructed by using two $\NS_-$ gates along with two beam splitters, but no
extra ancillae or measurements. Since both $\NS_-$ applications must succeed, the overall
probability of success for a $\CZ$ is $\frac{1}{16}$.
In order to make this construction scalable, one has to  
use {\em gate teleportation}~\cite{gottesman-chuang:1999}, which turns the gate
construction problem into a state preparation problem with the advantage
that such states can be prepared off line without {\em a priori} knowledge of the
inputs to the gate.
The idea of gate teleportation is to use the conjugation relations
of certain gates to modify the resource state and the correction operations of the
teleportation protocol, so that these gates can be applied implicitly
during the teleportation, in a manner similar to the protocol
for the $\CZ$ gate illustrated in Figure~\ref{fig:c-sign-tele}. Strictly speaking,
the $\CZ$ construction is based on the teleportation of some of the modes
constituting the qubits that must interact, and thus the operations are not
general unitaries but instead restricted to linear optics operations,
single photon sources, and post-selection based on single photon detection
of certain modes.

Even though the direct $\CZ$ construction has a small (but non-zero) 
probability of success, 
it is always known when the teleportation succeeds, and
the resource state of the teleportation is independent of the inputs to the operation
so one can make many attempts to produce such a resource state before performing
the teleportation. Alternatively, one may make many attempts in parallel, 
and select one that succeeds. While without teleportation the number of attempts
needed for error free quantum computation grows exponentially for
circuits constructed with the probabilistic $\CZ$ gates, 
 with the gate teleportation it grows linearly~\cite{klm:2001,klm-thr:2000}. 
Since the $\CZ$ along with all single
qubit unitaries form a universal set for quantum computation~\cite{nielsen-chuang:2001},
this gives an efficient construction for a universal quantum computer using only
linear optics elements.

Teleportation in linear optics cannot be performed deterministically~\cite{lutkenhaus:1999}, 
but it can be performed with probability arbitrarily close
to one at the cost of higher complexity for the teleportation protocol~\cite{klm:2001}. 
Once again, as in the probabilistic $\CZ$ construction, one can determine whether
the protocol succeeded through the measurement of ancillary modes, so possible
failures are always flagged by the measurement outcomes. Since it is not clear how
to increase the probability of success of the direct $\CZ$ construction, and
bounds for possible probabilities of success are predicted to be significantly
smaller than one~\cite{knill-bounds:2003}, this gate teleportation
scheme is highly necessary. We say that the probability of these intrinsic 
teleportation failures occurring is $\e_{\text{ideal}}$, since such failures
occur even when considering ideal photon detectors (i.e. with perfect efficiency)
and linear optics elements.

What makes the teleportation scheme highly attractive as well is exactly what
the failure entails. When one of the teleportations involved in the $\CZ$ fails (but no detectors fail),
the output is equivalent to a successful teleportation followed by a measurement
in the $\Z$ eigenbasis -- that is, we
can view it as a successful gate application followed by a $\Z$ measurement 
of one of the qubits independently with probability $\e_{\text{ideal}}$, and such an
event is automatically flagged by the outcome of the measurement of the ancilla modes. This is
what we call the {\em ideal hardware error model}, and it has been shown 
to be very benign, with an error threshold arbitrarily close to one~\cite{klm-thr:2000,manny:2001}.

Hence teleportation failures under this model mean
that one of the projectors into the eigenbasis of $\Z$ has been
applied. Exactly which projector was applied to the teleported qubit depends on the outcome of the 
measurement of the ancillae.
Formally, the projectors are $\Z_+=\frac{1}{2}\left(\I+\Z\right)$ and
$\Z_-=\frac{1}{2}\left(\I-\Z\right)$.
Since the location of these measurements is flagged by post-selection, this
type of failure is a form of {\em erasure} -- an error of known location.

\section{\label{sec:lossy-error-model}Lossy Hardware Error Model}

One of the largest technical hurdles in the implementation of these probabilistic
gates is the fact that
single photon detectors are notoriously difficult to build. While
for some wavelength very high efficiency can be obtained, the rate
at which false photon detections are signaled, the so called {\em dark counts},
is unacceptably high~\cite{sobolewski}. Dark counts are particularly
troublesome for the gates proposed in Ref.~\onlinecite{klm:2001} because
they could cause photon loss to go undetected, as well as causing
incorrect post-selection of states.
Recent proposals of photon detectors
based on phase transitions in superconductors have very low
dark count rates, 
although little attention has been given to 
optimizing the efficiency of these detectors. We are interested in finding
the minimum efficiency necessary for these detectors, given dark counts
stay negligible, in order to be able to perform useful quantum
computation~\footnote{In order to consider dark count induced errors, one would need
to look into a general error model, not an erasure error model.}.

Modifications to the gate teleportation protocol that allow for the detection of photon
loss at the detectors are known~\cite{klm:2001}. This protocol can differentiate
between the teleportation failures due to the limitations of linear optics (leaving 
the error model due to such failures intact) and the failures due to
photon loss at the detectors -- when this occurs, the corrupted qubit is replaced
with a fresh qubit in a known state.
We take the detection of a single physical qubit to fail with probability
$\dt$, and the overall probability of the $\CZ$ gate teleportation failing due to 
photon loss at the detectors to be $\e_{\text{loss}}$. The form of $\e_{\text{loss}}$ as a function of
$\dt$ depends on the choice of protocol used -- Ref.~\onlinecite{klm:2001}
describes a family of protocols -- and for the purposes of this paper, we
will take them to be independent parameters.
Once again, any possible error due to 
these types of failure is always flagged by the gate construction, which significantly
simplifies error correction since it is known {\em a priori} where the failures have
occurred. 

Consider the error model due {\em only} to photon loss at
the detectors ($\e_{\text{ideal}}=0$) -- such an error model is not physical, since
a linear optics quantum computer will always have teleportation failures, but
taking this limiting case simplifies the analysis significantly.
If we consider a single physical qubit measurement 
where photon loss occurred,  it is clear that all information about the 
qubit is lost. We can model this loss of information by the {\em full erasure} superoperator
\begin{equation}\label{eqn:full-erasure}
\er{E}(\rho)=\frac{1}{2}\I=\frac{1}{4}\left(\rho+\X\rho\X+\Z\rho\Z+\Y\rho\Y\right).
\end{equation}
However, because of the dual-rail encoding, it is always
clear which physical qubit measurement failed in such a manner
since at the lowest level of encoding qubit measurement consist of
measuring the two constituent modes, and thus only one of the detectors
may click. If a photon is not detected on either mode, a measurement failure 
has occurred, and we may replace the qubit with a fresh qubit in a known state. This is
fundamentally different from the depolarizing channel~\cite{bennett:97} where there is no
{\em a priori} knowledge of the position of the errors -- one may think of
this superoperator as the depolarizing channel superoperator conditioned
on perfect information about which qubits were randomized. 

It is clear that the full erasure does not commute with gate teleportation, otherwise
we could in principle transfer some of the information from the control
qubit of a $\CZ$ to the target even though all information was lost
in the failed teleportation of the control qubit.
We take a worst
case approach, and assume that {\em any} photon loss during one of the
teleportations yields {\em total} information loss of the qubit being teleported, 
and that the error model for the other qubit associated with the same gate
teleportation in which the photon was lost is determined by which correction
might have been needed to be applied. In the particular case of the linear optics proposal,
the error model is symmetric: if we disregard the classical correlation between the
errors of the outputs of the $\CZ$,  photon loss in one of the teleportations
translates to a full erasure of the qubit being teleported and a $\Z$ erasure
of the other qubit involved in the gate operation. This is because in both
cases the correction operation from one teleportation to the other is a $\Z$ gate, and if
it is not known whether such gate was supposed to be applied because
of the photon loss, we have the superoperator
\begin{equation}
\er{Z}(\rho)=\frac{1}{2}(\rho+\Z\rho\Z),
\end{equation}
which is what we call a {\em $\Z$ erasure}. Note that this superoperator
can be interpreted as an unintentional $\Z$ measurement of unknown outcome, since
\begin{equation}
\Z_+\rho\Z_++\Z_-\rho\Z_-=\frac{1}{2}\left(\rho+\Z\rho\Z\right)=\er{Z}(\rho),
\end{equation}
and thus this is fundamentally different from the 
phase erasure channel~\cite{bennett:97}, since it provides the added {\em a priori} knowledge of
when corruption has occurred or not.
As mentioned before, we say that there is a probability $\e_{\text{loss}}$ 
that photon loss occurs in one of
the teleportations in the $\CZ$ implementation, which entails a full erasure of the
qubit being teleported, and a $\Z$ erasure of the qubit to which it was coupled through
the $\CZ$ application. The correlation between the erasures on these two qubits
is ignored for our calculations, but could be exploited to obtain better thresholds.

In summary, all single qubit operations are taken to be error free, and
the $\CZ$ is taken to introduce at each output qubit either $\Z$ measurements, full erasures, or
$\Z$ erasures with finite probability. For simplicity, we
consider the different error models independently, that is, we calculate
the threshold for the case where only $\Z$ measurements occur (where the
hardware is ideal, with $\e_{\text{loss}}=0$ and $\dt=0$, but teleportation is imperfect, i.e.
$\e_{\text{ideal}}\not=0$), and we
calculate the threshold where only full and $\Z$ erasures occur (that is, where
the teleportation protocol is perfect, $\e_{\text{ideal}}=0$, but the detectors are not, so 
$\dt\not=0$ and $\e_{\text{loss}}\not=0$). 

A collection of single qubit erasures is referred to as an {\em erasure pattern}, and the
{\em weight} of the erasure pattern is the number of qubits that have been 
affected by an erasure, regardless of the type of erasure.

%%%%%%%%%%%%%%%%%%%%%%%%%%%%%%%%%%%%%%%%%%%%%%%%%%%%%%%%%%%%%%%%%%%%%%%%%
\section{\label{sec:error-correction}Fault-tolerance and Error Correction}

The quantum error correction code considered here to protect the data
from the error model in question is a $[[7,1,3]]$ self-orthogonal, doubly-even 
CSS code~\cite{bank-shor:1996,steane:1996} with stabilizer generators
\begin{equation}\label{eqn:steane-stabs}
\begin{array}{ccccccccccccccl}
\op{M_1}&=&\X&\ox&\X&\ox&\X&\ox&\X&\ox&\I&\ox&\I&\ox&\I\\
\op{M_2}&=&\X&\ox&\X&\ox&\I&\ox&\I&\ox&\X&\ox&\X&\ox&\I\\
\op{M_3}&=&\X&\ox&\I&\ox&\X&\ox&\I&\ox&\X&\ox&\I&\ox&\X\\ 
\op{M_4}&=&\Z&\ox&\Z&\ox&\Z&\ox&\Z&\ox&\I&\ox&\I&\ox&\I\\
\op{M_5}&=&\Z&\ox&\Z&\ox&\I&\ox&\I&\ox&\Z&\ox&\Z&\ox&\I\\
\op{M_6}&=&\Z&\ox&\I&\ox&\Z&\ox&\I&\ox&\Z&\ox&\I&\ox&\Z.
\end{array}
\end{equation}
Self-orthogonal CSS codes are particularly suited for the error 
model considered here because they yield 
simple constructions of fault-tolerant encoded Clifford group
operations. 
Although the Clifford group is not a universal set of quantum operations, it is
well known how to extend it in order to obtain a universal set 
fault-tolerantly~\cite{shor:1996,zlc:2000},
as well as how this affects the threshold value~\cite{gottesman-thesis:1997}.
In the case of the error models considered here, the threshold is unaffected
since we require that computation be performed only on error free states -- that is,
error correction is performed until an uncorrectable error occurs, causing
the computation to be aborted, or until the data is error free, at which point the 
computation may continue.

Recall that the Clifford group consists of all operations that preserve the
Pauli group under conjugation, and this group is generated by
the $\CZ$ gate, the Hadamard gate 
$\op{H}=\frac{1}{\sqrt{2}}\left(\begin{smallmatrix}1 & 1\\ 1 & -1\end{smallmatrix}\right)$,
and the phase gate $\op{P}=\left(\begin{smallmatrix}1 & 0\\ 0 & i\end{smallmatrix}\right)$. 
The $\CZ$ can be implemented transversally
by qubitwise $\CZ$s between two encoded qubits. Note that CSS
codes have a transversal encoded $\CZ$. 
If in addition $\op{H}$ has to be transversal, then the CSS code has to be
constructed from a single, self-orthogonal classical code. Moreover, if $\op{P}$
also has to be transversal, then all codewords in this classical code must have
doubly-even weight.
Since the 7-qubit code considered is both doubly-even and constructed 
from a self-orthogonal classical code,
the Hadamard and phase gates can be implemented 
using only single qubit operations qubitwise,
and therefore are error free under the models considered here. 
In order to obtain a universal gate set, we can add a non-Clifford gate such as 
the $\pi/8$ gate, which has a known
fault-tolerant construction~\cite{zlc:2000}. We will make use of the properties of doubly-even
CSS codes for the error correction circuits studied in Section~\ref{sec:error-correction-circuits}.

%%%%%%%%%%%%%%%%%%%%%%%%%%%%%%%%%%%%%%%%%%%%%%%%%%%%%%%%%%%%%%%%%%%%%%%%%
\section{\label{sec:encoded-error}Encoded Error Model}
Error models consisting of erasures yield particularly simple encoded error
models because it is always known when an error is unrecoverable. In such a case
one can simply take the code block to be an encoded erasure at the next level 
of encoding. In general, an uncorrectable failure is not an encoded failure,
and it requires further processing to make it an encoded failure, as will
be discussed below. What remains to be determined is what kind of encoded failure
results.

In the case of ideal hardware, all erasures are $\Z$ measurements of known
outcomes. Turning to the 7 qubit code described above, we have that all
weight one and two erasure patterns are correctable.
Of the $35$ possible weight three erasure patterns, 
 $28$ are correctable. The remaining $7$ weight three patterns, along with all other
patterns with higher weight~\footnote{There are weight four patterns that are correctable,
but we ignore them for simplicity, since the error recovery operation used here is not
applicable to these patterns.}, can be identified with an encoded $\Z$ measurement,
and are therefore uncorrectable failures. These collections of individual
single qubit measurements are not, strictly speaking, equivalent to an encoded
measurement, since they collapse the seven qubits into a state outside the
code space. However, because the encoded $\ket{\logic{0}}$ and $\ket{\logic{1}}$
are superpositions over mutually exclusive sets of states, it is easy to infer which
encoded state the measurement results correspond to, and then replace the qubits
with a fresh encoded $\ket{\logic{0}}$ or $\ket{\logic{1}}$. This operation is
taken to be error free, since we assume that state preparation can be attempted until 
no errors have occurred. Since all encoded failures are $\Z$ measurements, in the ideal
hardware error model the break even condition between encoded $\Z$ measurements
at the first level of encoding, denoted by $\Z^{(1)}$, and single qubit $\Z$ measurements,
$\Pr(\Z^{(1)}\text{ measurement})=\Pr(\Z\text{ measurement})$, implies
\begin{equation}
\e_{\text{ideal}}^{(1)}=\e_{\text{ideal}},
\end{equation}
where $\e_{\text{ideal}}^{(1)}$ is the failure rate at the first level of encoding.

In the case of the lossy error model, because each qubit can suffer either $\Z$ erasures
or full erasures, different uncorrectable errors on the $7$ qubit code 
will lead to different encoded errors. For simplicity we can take all encoded 
failures to be encoded full erasures,
which in the first level of encoding we denote $\er{E}^{(1)}$, so that
the break even condition on the probabilities $\Pr(\er{E}^{(1)})=\Pr(\er{E})$
implies
\begin{equation}
\e_{\text{loss}}^{(1)}=\frac{1}{2}\e_{\text{loss}},
\end{equation}
since only half of the lossy error model failures are full erasures, where
$\e_{\text{loss}}^{(1)}$ is the failure rate at the first level of encoding.
The resulting threshold is at most as high as the real threshold, considering 
the different kinds of
encoded erasures that are simpler to correct, but should be lower in general. 
A more detailed analysis can be made~\cite{msilva-thesis:2004,msilva:2004}, 
and exact probabilities distributions for the different kinds of erasures can be 
calculated using the same technique used to calculate the 
threshold in Section~\ref{sec:thresholds},
but the simplifying assumption made here is enough to 
match the prediction in Ref.~\onlinecite{klm-thr:2000}. 

In reality we would like to consider an error model that takes
both these sources of error into account. Section~\ref{sec:error-correction}
describes how each of the different kinds of erasures are corrected, and demonstrates
the progressively higher cost of correcting a $\Z$ measurement, a $\Z$ erasure and a full
erasure. Given this fact, for the error correction code chosen here, and for the error correction
technique used here, the threshold for an error model consisting of both types of failures
is bounded above by the
ideal error model threshold, and below by the lossy error model threshold.

%%%%%%%%%%%%%%%%%%%%%%%%%%%%%%%%%%%%%%%%%%%%%%%%%%%%%%%%%%%%%%%%%%%%%%%%%
\section{\label{sec:error-correction-circuits}Error Correction Circuits}
In general, in order to correct errors by using a stabilizer code, one
simply needs to measure the stabilizer generators and infer the most
probable error that occurred and apply the correction. In the case of
erasure errors, the knowledge of which qubits have been affected by the
error superoperator greatly reduces the number of stabilizer operators that
need to be measured. This is because we need only measure
stabilizer operators that act non-trivially on the qubits affected
by the error superoperators, and this greatly reduces the probability
of introducing more errors into the data.

\begin{figure}
\includegraphics[width=3.75cm]{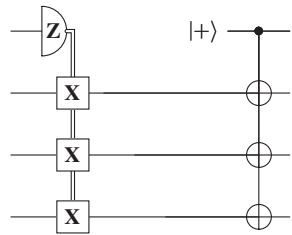}
\caption{\label{fig:z-erasure}A circuit for correcting a single
$\Z$ measurement in the top qubit as long as the remaining 3 qubits are
erasure-free. The four qubits are assumed to be part of a block
encoded with the 7 qubit code. The state $\ket{+}$ is
the $+1$ eigenvalue eigenvector of $\X$. The double lines
indicate control based on the outcome of the measurement.}
\end{figure}

\begin{figure}
\begin{center}
\includegraphics[width=7cm]{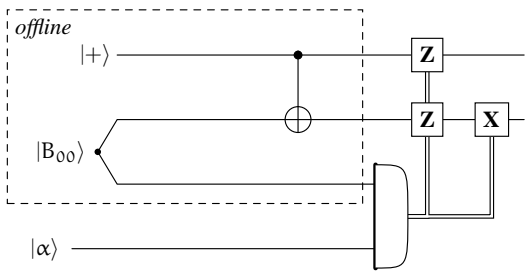}
\end{center}
\caption{Performing a $\CNOT$ between $\ket{+}$ and an arbitrary state $\ket{\alpha}$ through
teleportation.\label{fig:c-not-part-teleport}}
\end{figure}

This procedure can be further optimized by incorporating the syndrome
measurement and correction into a single step, as first described in
Ref.~\onlinecite{klm-thr:2000} for the case of $\Z$ measurement correction codes. 

The fact that the 7 qubit code employed here is based on a classical 
doubly-even code, allows us to consider the 4 qubit subsystem in 
the support of any given stabilizer operator. 
Since this is a CSS code, we can focus on
stabilizer operators that are made up of tensor products of $\X$s and
identities, and stabilizer operators that are made up of tensor products 
of $\Z$s and identities. 

Since we are only interested in correcting erasures of weight up to 3, 
we can consider a single qubit that has undergone 
some erasure, along with three qubits that are still intact. We choose stabilizer
operators that act non-trivially on all of these four qubits and trivially on
all other qubits -- this is always possible in the 7 qubit code.
Considering only this 4 qubit subsystem, the stabilizer operators in question 
are~\footnote{Note that this is different from the 4 qubit erasure code described
by Grassl and collaborators~\cite{grassl-etal:1997}, because for any 4 qubits we consider, there
will be other stabilizer operators in the 7 qubit code that are not generated by
these two.}
\begin{equation}\label{eqn:grassl-stabs}
\begin{array}{ccccccccl}
\op{M_1^{'}}&=&\X&\ox&\X&\ox&\X&\ox&\X\\
\op{M_2^{'}}&=&\Z&\ox&\Z&\ox&\Z&\ox&\Z.
\end{array}
\end{equation}
The usual approach is to measure these two operators fault-tolerantly
in order to determine what kind of Pauli correction needs to be applied
to the erased qubit. If we consider $\Z$ erasures, we need only
measure $\op{M_1^{'}}$. Alternatively, one can
simply use the circuit depicted in Figure~\ref{fig:z-erasure} (without 
loss of generality, we consider the corrupted qubit to be the 
first qubit of the four)~\footnote{To see how this circuit works, simply
consider the action of the circuit on the states stabilized by
\eqref{eqn:grassl-stabs}.}.
As demonstrated before, a code that can
correct $\Z$ erasures can also correct $\Z$ measurements at known locations, 
since a $\Z$ erasure can be given by a 
$\Z$ measurement of unknown outcome.
In this case, we do not need to perform the explicit measurement in the 
$\Z$ eigenbasis since we already have that information, but the rest of the
circuit remains as in Figure~\ref{fig:z-erasure} -- thus there is an added
cost of a possible measurement failure when correcting $\Z$ erasures. 
If the measurement does fail and the
qubit is destroyed (as is the case when a photon is detected),
one simply abandons the attempt at correcting the $\Z$ erasure, since it will
then be replaced by a full erasure. 

\begin{figure}
\begin{center}
\includegraphics[width=7cm]{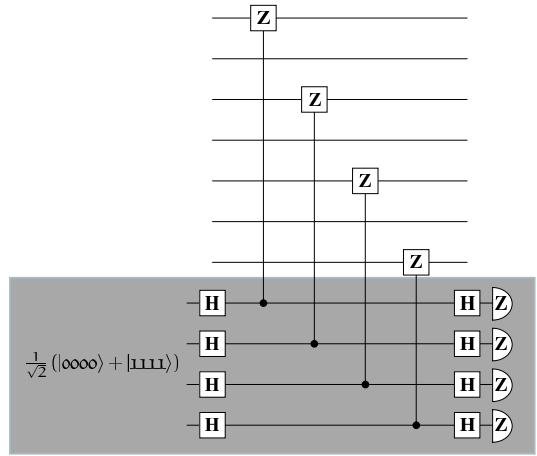}
\end{center}
\caption{Fault-tolerant measurement of stabilizer operator $\Z\ox\I\ox\Z\ox\I\ox\Z\ox\I\ox\Z$
as proposed by Shor~\cite{shor:1996}. The qubits in the shaded region are ancillas, 
and the measurement of the operator is inferred from the
measurement of the 4 ancilla qubits.\label{fig:ft-stab-meas}}
\end{figure}

The circuit in Figure~\ref{fig:z-erasure} is not fault-tolerant, as it stands.
In order to make it fault-tolerant, one simply applies a modified teleportation
protocol to the three error-free qubits~\cite{klm:2001,manny:2001,msilva-thesis:2004}. 
In order to understand why this is fault-tolerant, 
consider the teleportation of the three intact qubits
before applying the $\CNOT$s in Figure~\ref{fig:z-erasure}. Since $\CNOT$ is a
Clifford gate, we can apply
the $\CNOT$s to the Bell states needed for teleportation, and simply modify the
recovery stage of the teleportation, in a manner similar to what was described in
Figure~\ref{fig:c-sign-tele}.
Note that it is unnecessary to teleport the control qubit, i.e. the qubit in the
state $\ket{+}$, since it is a fixed resource state.
Unlike the $\CZ$ gate construction, where photon modes are teleported,
this is a teleportation of the qubits, and can be thought of in terms of 
the usual higher level gates such as $\CNOT$s and Pauli operators.

This procedure becomes clearer if we consider the teleportation of only one
of the qubits, followed by the $\CNOT$ with the resource state $\ket{+}$, 
and propagate the $\CNOT$ backwards, 
as described in Figure~\ref{fig:c-not-part-teleport}. If there is a failure
in the Bell measurement of the qubits, it will only possible cause a $\Z$
erasure in the control bit, as well as a possible full erasure on the 
target bit. Since $\Z$ errors do not propagate from the
control of the $\CNOT$ gate, the other two $\CNOT$s can be performed fault
tolerantly in parallel.

In summary, we simply need to measure
the erased qubit in the $\Z$ eigenbasis, discard the erased bit, and apply
the teleportation-based fault-tolerant version of Figure~\ref{fig:z-erasure}
to three erasure free qubits in the codeword and an extra qubit in the $\ket{+}$ state
to replace the discarded one, and the $\Z$ erasure will be corrected. 
If any of the teleportations fail, the failure will affect only the qubit
we are attempting to recover and the qubit that was being teleported.

The circuit in Figure~\ref{fig:ft-stab-meas} is used to measure stabilizer operators
made of tensor products of $\Z$s and $\I$s, and thus it partially corrects full erasures,
yielding a $\Z$ erasure if successful. The fault-tolerant version of the
circuit in Figure~\ref{fig:z-erasure} can be used to correct $\Z$ erasures as well
as unintentional $\Z$ measurements. The strategy taken is to correct full erasures first,
and once there are no more full erasures, to correct $\Z$ erasures and $\Z$ measurements.
We do not claim that this strategy and circuits are optimal for error correction, 
but this choice simplifies the exact calculation of the threshold significantly. 

%%%%%%%%%%%%%%%%%%%%%%%%%%%%%%%%%%%%%%%%%%%%%%%%%%%%%%%%%%%%%%%%%%%%%%%%%
\section{\label{sec:thresholds}Thresholds}

The {\em threshold theorem}~\cite{aharonov,klw,preskill} 
states that by concatenated coding -- that is, the repeated encoding
of a quantum state -- one can perform quantum computation with arbitrarily small error
efficiently as
long as the error is below a certain threshold. We take this threshold to
be the smallest probability of error such that the probability of 
an encoded failure $\e^{(1)}$ is equal to the probability of a single unencoded 
qubit failure $\e^{(0)}\equiv\e$.

The encoded error rate can be calculated by tracking the probability
of going from any given erasure pattern to any other erasure pattern
during an attempt at error correction. This describes a Markov chain,
and such a description can be made more compact by considering symmetries
of the error correction code and of the error correction circuitry
~\cite{msilva-thesis:2004}. Erasure patterns can be grouped
into equivalence classes defined by the error correcting code as well
as the error correcting procedure, and we need only consider probabilities 
of going from one equivalence class to another -- in the case of the 7 
qubit code, we need to consider only $11$ equivalence classes versus
the $128$ that would be necessary for a naive description of the Markov 
chain. It is straightforward to obtain the initial distribution of the 
different equivalence classes as well as the transition matrix of the Markov
chain that describes the 
change in the distribution due to one error correction attempt. The distribution
after multiple error correction attempts can be obtained by taking higher
powers of the transition matrix and applying it to the initial distribution.
Each of the non-trivial equivalence classes of erasure patterns 
can be associated with an encoded erasure, and therefore one can obtain
the error distribution at any given encoding level. In some cases,
there is additional processing and transitions between different
equivalence classes associated with the mapping between the erasure pattern
and an encoded erasure, since, for example the erasure operator
$\openone\otimes\openone\otimes\openone\otimes\openone\otimes\openone\otimes\openone\otimes\er{E}$
is not equivalent to any encoded operation. In order to account for such
processing, another transition matrix would be required.
Details of this procedure are discussed elsewhere~\cite{msilva:2004}.
In the case of the ideal hardware error model, all erasure patterns
can be mapped directly to encode measurements by measuring all the
qubits of a code block that is not erasure free. In the case of
the lossy hardware model, we take a worst case approach and all
erasure patterns at the end of the error correction procedure
are taken to be an encoded full erasure, so there is no
need for the more detailed analysis -- we can simply replace
the corrupted block of qubits with a block in a known fixed encoded state.

\begin{figure}
\begin{center}
\includegraphics[width=7cm]{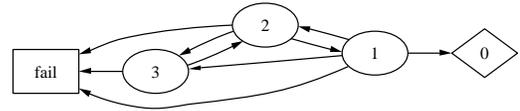}
\end{center}
\caption{The Markov chain describing the error recovery for the ideal hardware error
model. The states are labeled by the weight of the correctable erasures, and the
'fail' corresponds to all patterns that are not correctable according to our procedure. 
Probabilites are omitted for readability.\label{fig:z-graph}}
\end{figure}

In the case of perfect hardware ($\dt=0$), 
the error rate recursion relation is
\begin{equation}
\e_{\text{ideal}}^{(1)}=56\e_{\text{ideal}}^3+406\e_{\text{ideal}}^4+3878\e_{\text{ideal}}^5-129675\e_{\text{ideal}}^6+\cdots,
\end{equation}
which yields a threshold, for $\e_{\text{ideal}}^{(1)}=\e_{\text{ideal}}$, of approximately
$\e_{\text{ideal}}=0.115$. The Markov chain describing the error recovery
procedure for this model is shown in Figure~\ref{fig:z-graph}.

\begin{figure}
\begin{center}
\includegraphics[width=6cm,angle=270]{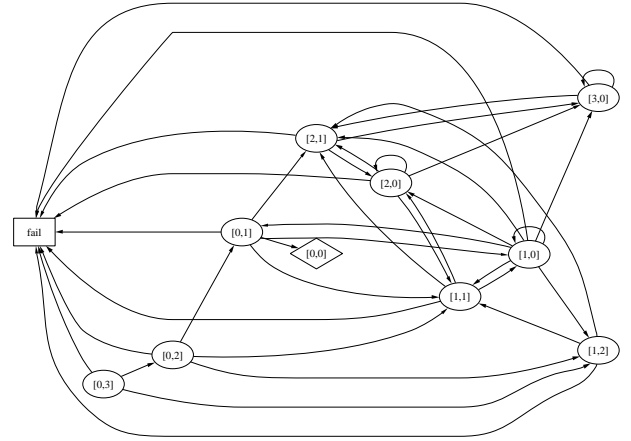}
\end{center}
\caption{The Markov chain describing the error recovery for the lossy hardware error
model. The states are labeled $[m,n]$ according to the number of full 
erasures $m$ and the number of $\Z$ erasures $n$, and the
'fail' state corresponds to all patterns that are not correctable according to our procedure. 
Probabilites are omitted for readability.\label{fig:full-graph}}
\end{figure}

In the case of lossy hardware with perfect teleportation, 
assuming that $\dt=\e_{\text{loss}}$, the error rate recursion relation is
\begin{equation}
\e_{\text{loss}}^{(1)}=1050\e_{\text{loss}}^{3}+33173\e_{\text{loss}}^4-46242\e_{\text{loss}}^5-6861701\e_{\text{loss}}^6+\cdots,
\end{equation}
which yields a threshold, for $\e_{\text{loss}}^{(1)}=\frac{1}{2}\e_{\text{loss}}$, of
approximately $\e_{\text{loss}}=0.0178$. This threshold is only valid if
it is identical to or smaller than the encoded measurement threshold. Because
of the structure of CSS codes, encoded basis states correspond to superpositions of 
elements of a coset of a linear classical code. In the case of the $[[7,1,3]]$ quantum code,
the linear classical code is a $[7,4,3]$ code. Measurement failures 
can then be seen as classical erasures on this $[7,4,3]$ code, and ignoring
the correctable classical erasure patterns of weight three and higher, the encoded failure
rate for measurements is given by
\begin{equation}
\dt^{(1)}=\sum_{i=3}^7{7 \choose i}\dt^i(1-\dt)^{7-i},
\end{equation}
which yields the benign error threshold $\dt=0.25$, validating
the calculated threshold value $\e_{\text{loss}}=0.0178$ under the
assumption $\dt=\e_{\text{loss}}$. This is a worst case assumption because the
family of teleportation protocols described in Ref.~\onlinecite{klm:2001}
use an increasing number of detectors to increase the probability of
success, and for the smallest such protocol the probability of
photon loss is the same as the probability of photon loss for a
single qubit measurement. The Markov chain for the lossy hardware error model
is depicted in Figure~\ref{fig:full-graph}.

%%%%%%%%%%%%%%%%%%%%%%%%%%%%%%%%%%%%%%%%%%%%%%%%%%%%%%%%%%%%%%%%%%%%%%%%%
\section{\label{sec:conclusions}Conclusions}

Using the error correction techniques outline here, the error threshold for
Clifford gates is found to be at least $0.0178~<~\e~<~0.115$ (since
$\e_{\text{loss}}<\e<\e_{\text{ideal}}$), where $\e$ is the probability
that some type of erasure is introduced due to photon loss at the detectors 
or due to a teleportation failure.

The threshold values calculated here can be improved by using optimized
stabilizer measurement techniques~\cite{steane-filter:2002,steane-css-networks:2003}, 
or by merging stabilizer measurement and error correction steps 
more aggressively~\cite{manny:2001,knill:2003}. Figure~\ref{fig:ft-stab-meas} is a 
straightforward generalization of a technique used for a two qubit $\Z$ measurement
error correcting code~\cite{klm-thr:2000}, but Knill showed that by merging all
stabilizer measurements with error correction steps in modified teleportation 
protocols, significantly higher thresholds can be obtained~\cite{knill:2003}. 

The Markov chain description of the error correction procedure, discussed in more
detail elsewhere~\cite{msilva:2004}, can be used with any of these techniques.
This systematic approach to the calculation of the encoded error rates is particularly
useful for practical applications of concatenated codes, since it is able
to give the probability distribution of the encoded errors. The calculation to check
the dependency of the distribution on parameters such as the number of error correction
attempts or the number of concatenation levels is straightforward, and one could determine
easily how many correction attempts or concatenation levels are necessary to obtain some
target error rate.

%%%%%%%%%%%%%%%%%%%%%%%%%%%%%%%%%%%%%%%%%%%%%%%%%%%%%%%%%%%%%%%%%%%%%%%%%
\begin{acknowledgments}
M.S. would like to thank Michele Mosca and Raymond Laflamme for their support. 
This work was carried out while M.R. and C.Z. were at the Institute for 
Quantum Computing, University of Waterloo.
This research was partially funded by NSERC, MITACS and ARDA.
\end{acknowledgments}

%%%%%%%%%%%%%%%%%%%%%%%%%%%%%%%%%%%%%%%%%%%%%%%%%%%%%%%%%%%%%%%%%%%%%%%%%
\bibliography{bib-eloqc-thr}% Produces the bibliography via BibTeX.

\end{document}